# Design of 32-channel TDC Based on Single FPGA for μSR Spectrometer at CSNS

Fanshui Deng, Hao Liang, Bangjiao Ye, and Jingyu Tang

*Abstract*—**Muon Spin Rotation, Relaxation and Resonance (uSR) technology has an irreplaceable role in studying the microstructure and properties of materials, especially micro-magnetic properties. An experimental muon source is being built in China Spallation Neutron Source (CSNS) now. At the same time, a 128-channel uSR spectrometer as China's first uSR spectrometer is being developed. The time spectrum of uSR can be obtained by fitting the curve of positron count rate with time. This paper presents a 32-channel Time-to-Digital Converter (TDC) implemented in a Xilinx Virtex-6 Field Programmable Gate Array (FPGA) for measuring the positron's flight time of μSR Spectrometer. Signal of each channel is sampled by 16 equidistant shifted-phase 200 MHz sampling clocks, so the TDC bin size is 312.5ps. The measuring range is up to 327us. This TDC has the ability to store multiple hit signals in a short time with a deep hit-buffer up to 512. Time tag is added to each data to record the moment when the data was detected. Programmable time window and channel shielding give the flexibility to choose the time range and channels of interest. The delay of each channel can be calibrated. The data is transmitted to data acquisition system (DAQ) through Gigabit Ethernet. TDC and control logic are configured in real time by DAQ. The results of test show that the Full Width at Half Maximum (FWHM) precision of single channel is better than 273 ps with a low sensitivity to temperature and the linearity is pretty well.**

*Index Terms*—**TDC, shifted clock sampling, FPGA, μSR, CSNS.**

## I. Introduction

THE use of muon's spin properties to study the properties of materials is called Muon Spin Rotation, Relaxation and Resonance (μSR) technology. μSR technology has an irreplaceable role in studying the microstructure and properties of materials, especially micro-magnetic properties. Currently only ISIS in the UK, J-PARC in Japan, PSI in Switzerland and TRIUMF in Canada have high-performance muon sources for μSR applications. An experimental muon source is being built in China Spallation Neutron Source (CSNS) now [1], [2]. At the same time, a 128-channel μSR spectrometer as China's first μSR spectrometer is being developed.

This μSR spectrometer uses a double ring structure with 64 detectors placed in front of and behind the sample. The moment that muon beam enters the sample chamber serves as a start signal, and the moment that the detector around the sample detects positrons generated by muon decay serves as a stop signal. Muon's lifetime can be obtained by measuring the positron's flight time. The time spectrum of μSR can be obtained by fitting the curve of positron count rate with time and then can reveal the internal physical structure of the sample information. So the measurement of positron's flight time is the key to the entire μSR spectrometer. In order to measure the positron's flight time, a 32-channel Time-to-Digital Converter (TDC) is implemented in a Xilinx Virtex-6 Field Programmable Gate Array (FPGA).

## II. Design and Implementation

### A. *Design Method*

TDC design based on FPGA has two main methods [3]. The one is delayed data sampling (DDS), the input signal is routed to a number of delay units connected in series, and each delayed signal is sampled by the flip-flop using a common clock. The other one is shifted-phase clock sampling (SCS), the input signal is sampled with flip-flops using a set of equidistant shifted-phase sampling clock. So far, high-precision TDC has been implemented in a single FPGA using dedicated carry chains [4]. But the price is the consumption of a large number of logical resources. Taking into account the precision requirements in this project is not high and a great deal of control logic is needed, so this TDC uses the second method.

Xilinx virtex-6 FPGA has 32 global clock lines but only 12 global clock lines can be routed in one region. In this design, the external clock is multiplied by two PLLs to generate eight 200MHz clocks that have the same frequency but a difference of 22.5 degrees between the two adjacent clocks. Because flip-flops in FPGA can be configured to be activated by the rising edge or falling edge of the clock, the input signal can be routed to 16 flip-flops, with 8 flip-flops being activated by 8 clock rising edge and the other 8 flip-flops being activated by 8 clock

Manuscript received June 15, 2018. This work was supported by the National Natural Science Foundation of China under Grants 11527811.

F. S. Deng is with the State Key Laboratory of Particle Detection and Electronics, University of Science and Technology of China (USTC), Hefei 230026, China (e-mail: dfsh@mail.ustc.edu.cn).

H. Liang is with the State Key Laboratory of Particle Detection and Electronics, University of Science and Technology of China (USTC), Hefei 230026, China (e-mail: simonlh@ustc.edu.cn).

B. J. Ye is with the State Key Laboratory of Particle Detection and Electronics, University of Science and Technology of China (USTC), Hefei 230026, China (e-mail: bjye@ustc.edu.cn).

J. Y. Tang is with Institute of High Energy Physics, Chinese Academy of Sciences, Beijing 100049, China (e-mail: tangjy@ihep.ac.cn).

falling edge. This is equivalent to the input signal being sampled by 16 equidistant shifted-phase 200MHz clocks. The TDC bin size is Tclk/N, where Tclk is the period of sampling clock, and N is the number of sampling clocks. So the bin size is 5ns/16=312.5ps.

The precision of TDC is greatly limited by the linearity of TDC bins, and the most significant effect on the linearity of TDC bins is the time skew of input signal routing to 16 flip-flops. In order to minimize time skew, the LUTs (Look-Up Tables) in FPGA are used as shown in Fig. 1. The input signal is routed to 16 flip-flops after four stages of LUTs and these LUTs are placed symmetrically in the same slice column [5]. The INIT parameter for the FPGA LUT primitive is what gives the LUT its logical value. Set the INIT value to 2'h2, then when the LUT input is 0, its output will be configured to 0, when the LUT input is 1, its output will be configured to 1. Thus the output of the LUT will be equal to the input. After sufficient area constraints and timing constraints, the time skew can be reduced to a few picoseconds.

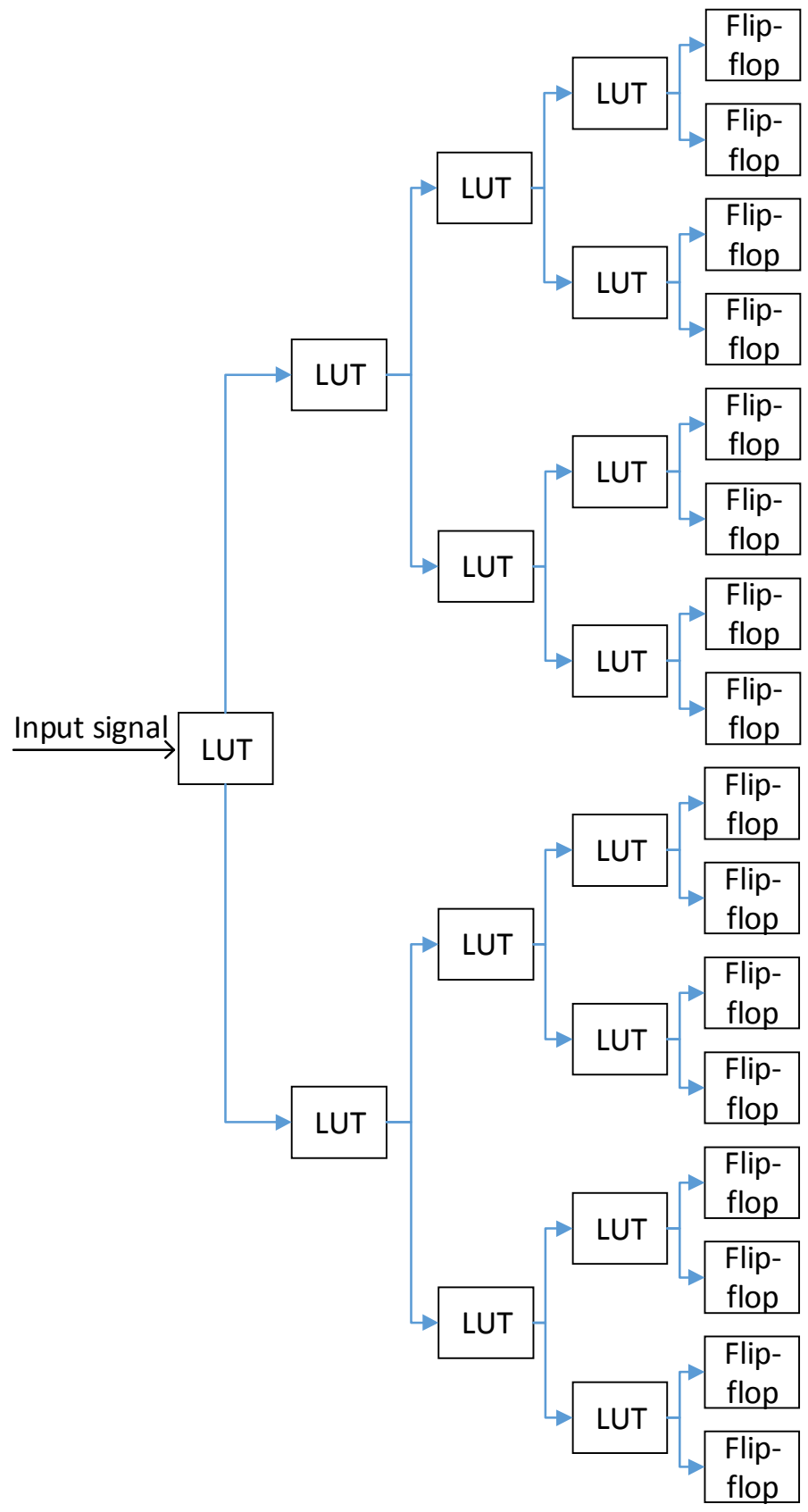

Fig. 1. Time skew is minimized by inserting LUTs.

## B. TDC Design

Fig. 2 is the schematic of one TDC channel. The input signal is routed to 16 flip-flops driven by 16 shifted-phase clocks with minimal time skew, the value of flip-flops is encoded into a 4-bit binary code as a fine count and then the fine count is sent to the hit-buffer with coarse count. The data is finally stored in the readout FIFO after passing through the trigger counter. Because the time measurement range is required to reach hundreds of microseconds, the coarse counter is a 16-bit counter driven by 200MHz clock, so the time measurement range is up to 327us. After the arrival of the Muon pulsed beam, the single-channel detector may receive multiple positrons in a short time, and these signals need to be recorded, so the TDC is designed to be multi-stop and it has a deep hit-buffer up to 512. The trigger counter can be configured as external trigger or internal self-trigger to decide whether the data is reserved or discarded and to record the number of trigger signals. The time tag with second level is used to record the moment when the hit signal is detected.

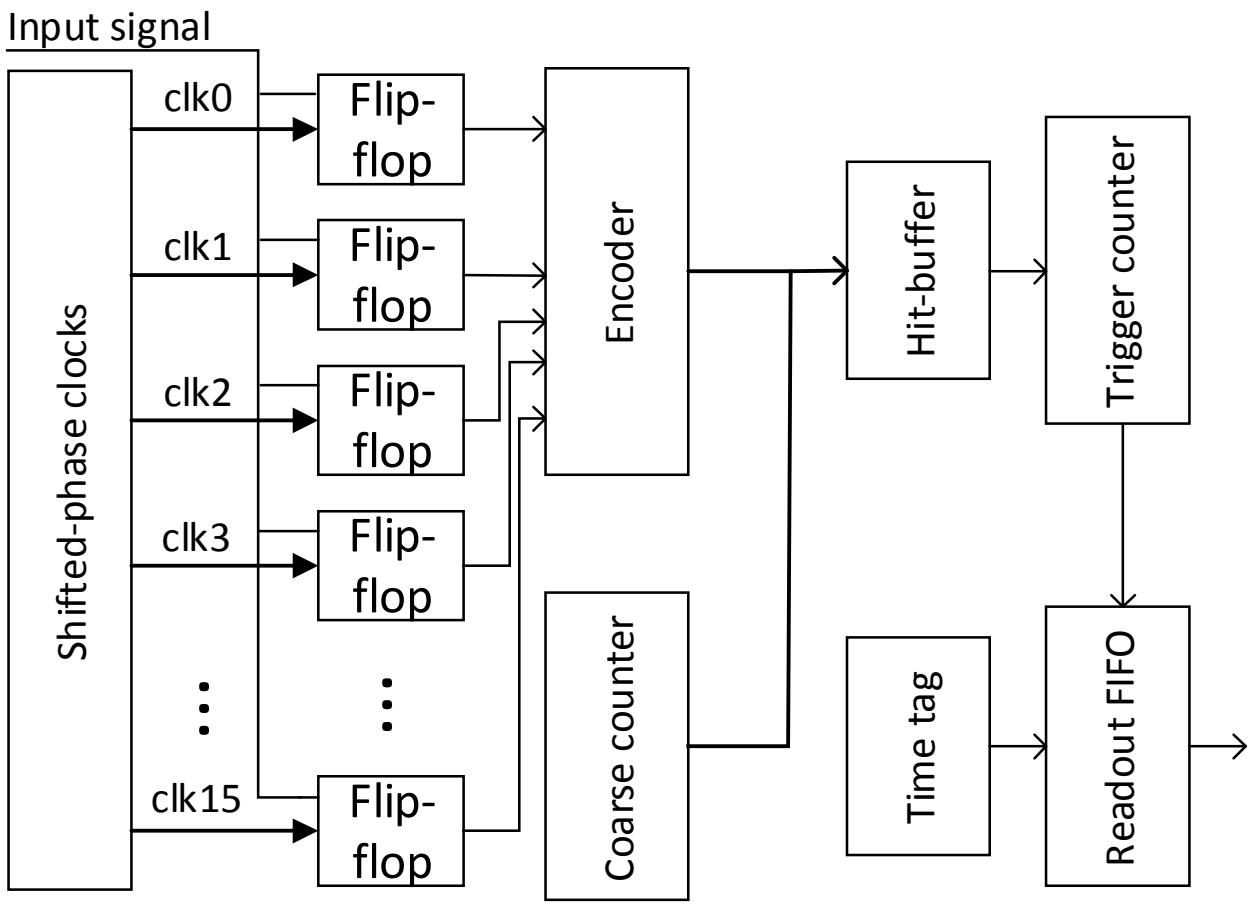

Fig. 2. Schematic of one TDC channel.

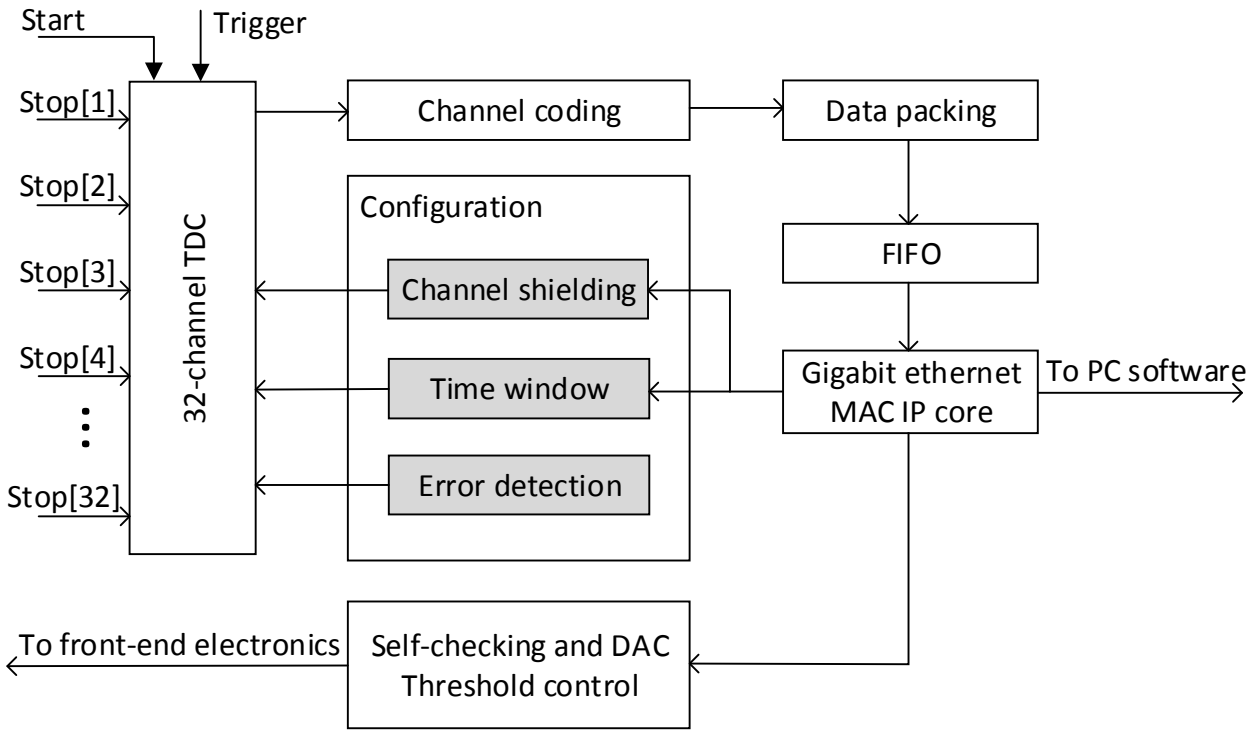

Fig. 3. Schematic of 32-channel TDC and control logic in a single FPGA.

## C. FPGA Implementation

This TDC contains a start channel and 32 stop channels, so only one coarse counter is needed. Schematic of 32-channel TDC and control logic in a single FPGA is shown in Fig. 3. 54bit data of each TDC, including 20bit time data, 26bit time tag and 8bit trigger count, is packaged into 64bit after channel coding and then stored in the FIFO, and finally uploaded to data acquisition system (DAQ) through Gigabit Ethernet. Because the entire system currently has 128 channels, a 7-bit channel coding is used as the identification number for each channel. TDC's configuration section includes channel shielding, time window setting and error detection. Channel shielding and time window are used to select the channel and time range of interest to the user. Error detection is used to reset the entire TDC logic when an error occurs.

This FPGA is also responsible for controlling the front-end electronics, including the self-checking and Digital to Analog Converter (DAC) threshold for each channel. The front-end electronics periodically simulate the detector signals and the FPGA also generates periodic pulse signals when self-checking. The pulse signal generated by the FPGA is sent to the TDC serves as a start signal and the timing pulse outputs by the front-end electronics are also sent to the TDC serve as stop signals. On the one hand, the entire electronics system can be tested whether it can work normally without the detector connected, on the other hand, the delay of each channel can be calibrated. TDC and control logic can be configured in real time by DAQ.

## III. Test Results

We tested the time measurement performance between the start channel and one of the stop channels. The test method is to generate the periodic start signal and stop signal by the Tektronix AFG3252C dual channel arbitrary / function generator, start signal and stop signal are then connected to the TDC, and the time measurement results are uploaded to DAQ via Gigabit Ethernet. Test results are shown in Fig. 4, the FWHM precision is equal to 368.1ps.

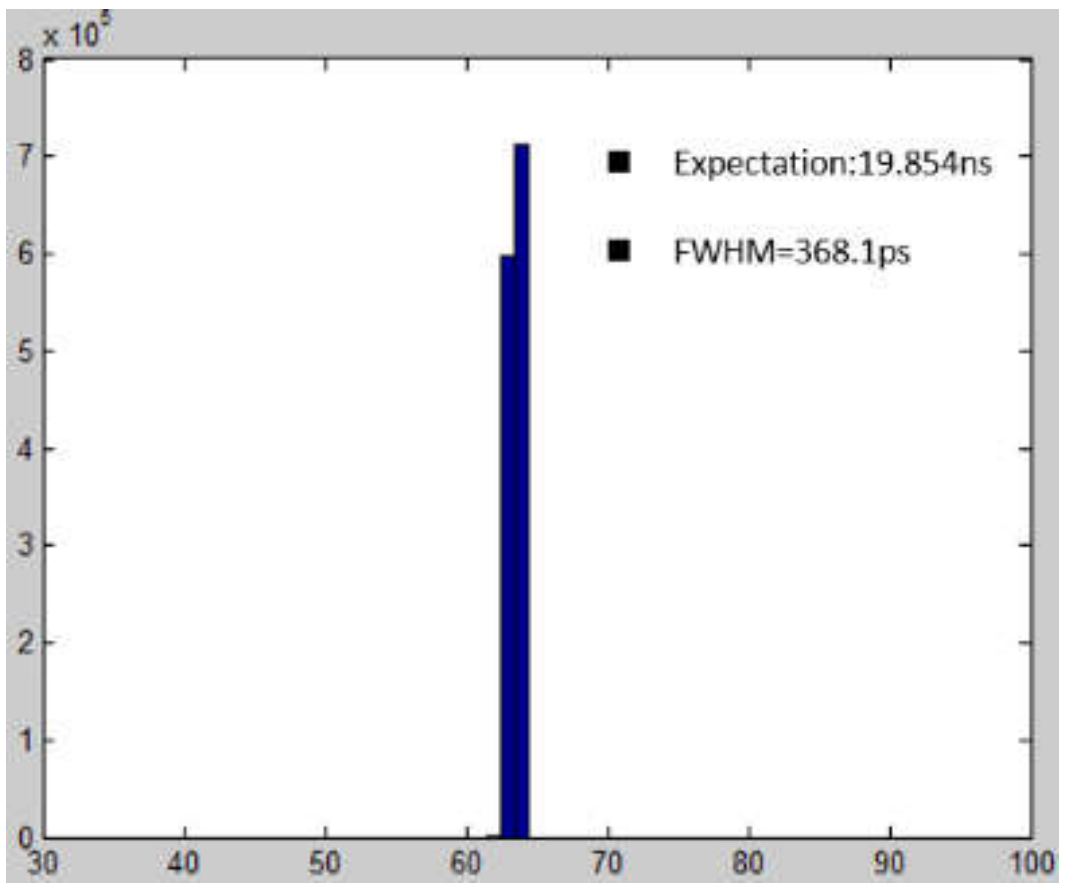

Fig. 4. Measured FWHM precision of TDC channel.

Change the time interval between the start signal and the stop signal, making multiple measurements in 200 ps steps. After 27 times measurements, the worst is 385.8ps. So the FWHM precision of single channel is better than 385.8/$\sqrt{2}$=273ps with a low sensitivity to temperature. The fitting curve of expectation with input time interval is shown in Fig. 5, the correlation coefficient R^2 is close to 1.0000 (0.99999989), so the linearity is pretty well.

## IV. Conclusion

In this paper, a 32-channel TDC is implemented in a single Xilinx Virtex-6 FPGA to measure the positron's flight time of μSR Spectrometer at CSNS. This TDC has the ability to store multiple hit signals in a short time and the measuring range is up to 327us. The FWHM precision is better than 273ps and the linearity is pretty.

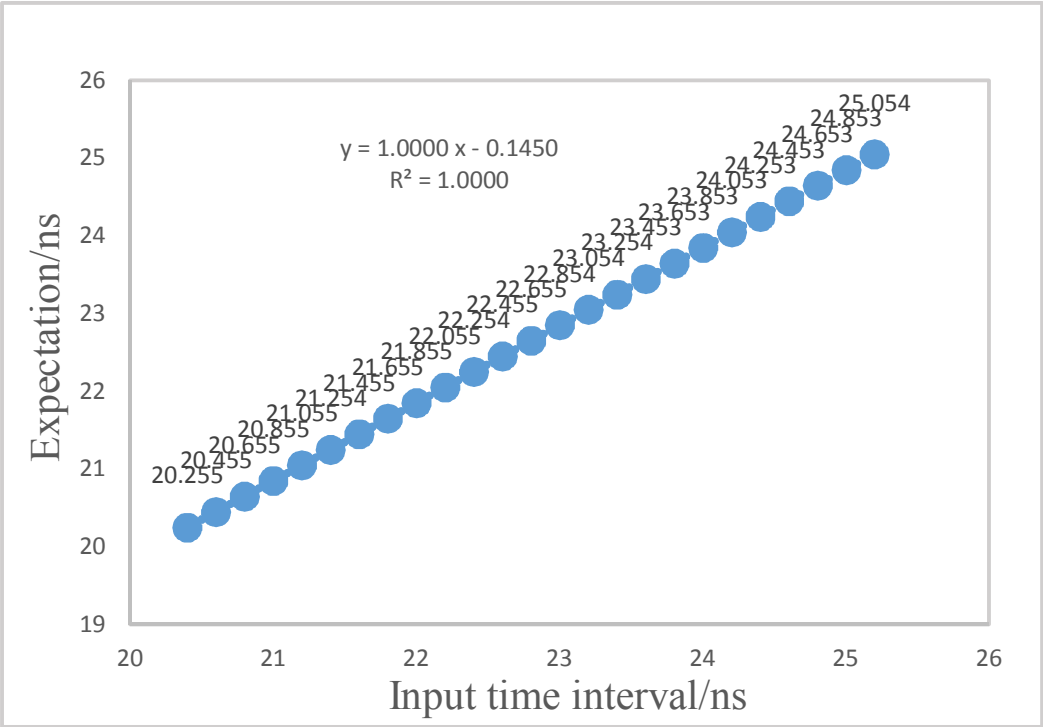

Fig. 5. Fitting curve of expectation with input time interval.

## References

[1] S. N. Fu, H. S. Chen, Y. W. Chen, Y. B. Chen, H. Y. Dong, S. X. Fang, K. X. Huang, W. Kang, J. Li, H. C. Liu, L. Ma, H. F. Ouyang, H. M. Qu, H. Sun, J. Y. Tang, C. H. Wang, Q. B. Wang, S. Wang, T. G. Xu, Z. X. Xu, C. Zhang, J. Zhang, " Status of csns project," Proceedings of the 4th International Particle Accelerator Conference, pp. 3995-3999, 2013.

[2] Y. Chen, H. T. Jing, J. Y Tang,"Physical Design of the Superferric Dipole for EMuS," IEEE Transactions on Applied Superconductivity, vol. 28, Apr. 2018.

[3] M. Büchele, H. Fischer, F. Herrmann, K. Königsmann, C. Schill, S. Schopferer, "The GANDALF 128-Channel Time-to-Digital Converter," Physics Procedia, vol. 37, pp. 1827-1834, 2012.

[4] J. Kuang, Y. G. Wang, Q. Cao, C. Liu, "Implementation of a high precision multi-measurement time-to-digital convertor on a Kintex-7 FPGA," Nuclear Instruments and Methods in Physics Research Section A: Accelerators, Spectrometers, Detectors and Associated Equipment, vol. 891, pp. 37-41, 2018.

[5] Y. G. Wang, P. Kuang, C. Liu, "A 256-channel multi-phase clock sampling-based time-to-digital converter implemented in a Kintex-7 FPGA," Conference Record - IEEE Instrumentation and Measurement Technology Conference, vol. 2016-July, Jul. 2016.